\documentclass{article}
\usepackage{graphicx} 
\usepackage{amsmath,amsfonts,amssymb,mathtools}
\usepackage{overpic}
\usepackage[
  backend=biber,
  style=numeric,      
  sorting=none,
  doi=true,
  eprint=true,        
  url=true,
  isbn=false
]{biblatex}
\newcommand{\A}{A}
\newcommand{\X}{X}
\newcommand{\Tr}{\mathrm{Tr}}

\title{Normal Matrix Model and Hyperelliptic Curves}
\author{Valdemar Melin \\ \small Department of Physics, KTH Royal Institute of Technology, Stockholm, Sweden}
\date{ October 21, 2026}

\begin{document}
\maketitle
\begin{abstract}
    We present a two-matrix model with a normal matrix with eigenvalues confined to contours in the upper half plane coupled to a Hermitian matrix. This generalizes the matrix model proposed by Jens Hoppe which corresponds to a single horizontal line. We solve the saddle point equations for several lines and show that the large-$N$ free energy can be expressed in terms of hyperelliptic integrals. We show that the model describes a Coulomb gas in the presence of an infinite conducting boundary.
\end{abstract}

\section{Introduction}
The large-$N$ solution of the two-matrix model with commutator-squared interaction, known as the Hoppe model, was first presented by Hoppe as a mathematical curiosity in his study of massless relativistic surfaces \cite{Hoppe:1982,Hoppe}. Since then, the significance of the model has been gradually uncovered, and today it is known to be related to holography, gravity, integrable models, and planar graph coloring \cite{Hoppe:1999xg,Kazakov:1998ji,Eynard:1998he,Murali:2026zse,Russo:2013qaa,Vescovi:2024fwt}.

In this paper, we introduce a generalization of the Hoppe model and its solution. As shown in a separate paper \cite{mwz}, this generalized model arises in the large-$N$ solution of certain Bethe ansatz-solvable models. As will be shown, the saddle point equation of this model describes a Coulomb gas in an external potential with an infinite conducting plate at the real axis. This generalized model is obtained by reinterpreting the Hoppe model as describing unitarily diagonalizable matrices with complex eigenvalues, i.e. normal matrices.

The model is formulated in terms of a Hermitian matrix $\A$ of size $N\times N$ and a normal matrix $\X$ of size $N\times N$ which has eigenvalues confined to contours $\gamma_a$ in the complex upper half plane.
Of particular importance will be the case when $\gamma_a$ are $n$ horizontal lines with imaginary parts $g_a>0$, i.e. $\gamma_a=\{z\in \mathbb C : \mathrm{Im}\,z = g_a\}$. In the space of such $\A$ and $\X$ we consider the model
\begin{align}
    \mathcal{Z}=\int e^{-N\Tr\left(\frac{1}{2i}(V(\X)-V(\X)^\dagger)-(\X \A-\A \X^\dagger)^2\right)}d\X\,d\A.\label{partfunc}
\end{align}
where $V$ is a polynomial with asymptotics at infinity making the integral convergent. For the Hoppe model and the case considered in section \ref{sec:4}, we take
\begin{align}
    V(z)=\frac{z^3}{3}. \label{cubic}
\end{align}
The measure $d\A$ is the standard translation invariant measure. For the measure $d\X$, let $dU$ be the $U(N)$ Haar measure and $z_1,\ldots, z_N$ the eigenvalues of $X$. The measure $d\X$ is defined as
\begin{align}
    d\X=dU\prod_{i<j}|z_i-z_j|^2  |dz_1|\cdots |dz_N|
\end{align}
where each $z_j$ belongs to a {\it fixed} contour $\gamma_{a_j}$ and $|dz|=|dz/dt|dt$ measures the arc length along the contours $\cup_a \gamma_a$, each contour being parametrized by a real parameter $t$. We are thus working in the "canonical ensemble" and when we take the limit $N\rightarrow \infty$ we will keep the fraction of eigenvalues belonging to a particular contour fixed, i.e. $N_a=P_aN$ where $N_a$ is the number of eigenvalues on the contour $\gamma_a$.
When $z_j\in\gamma_a=\{z\in \mathbb C : \mathrm{Im}\, z=g_a\}$ we have $|dz_j|=d\mathrm{Re}\,z_j$. The measure $d\A$ can be written the same way as $d\X$ by confining all eigenvalues to the real line.

We will be interested in observables of $\X$. 
The resolvent is given by
\begin{align}
    R_X(z)=\frac{1}{N}\left\langle\Tr\,\frac{1}{\X-z}\right\rangle
\end{align}
and the conjugate resolvent satisfies
\begin{align}
    R_{X^\dagger}(z) = \overline{R_X(\overline{z})}.
\end{align}
A key observable will be the combination
\begin{align}
    W(z)=V'(z)+2iR_{X}(z)-2iR_{X^\dagger} (z). \label{Wdef}
\end{align}
It has an asymptotic expansion (geometric series) at $z\rightarrow \infty$ of the form
\begin{align}
    W(z)=V'(z) + 2\sum_{m=1}^\infty \frac{\langle\Tr (\X^m)\rangle-\langle \Tr(\X^{\dagger m})\rangle}{iNz^{m+1}} \label{asymptoticW}
\end{align}
and thus is a generating function of the imaginary part of the trace moments $\mathrm{Im}\langle \Tr \X^m\rangle$. 

The connection to the Hoppe model is the case when $V$ is as in \eqref{cubic} and the eigenvalues of $\X$ lie on the line $\mathrm{Im}\,z =g$. We write $\X=Y+igI$ where $Y$ is Hermitian. Substituting this into the action in \eqref{partfunc}, we get
\begin{align}
    \Tr\left(\frac{\X^3-\X^{\dagger 3}}{6i}-(\X\A-\A\X^\dagger)^2\right)=\Tr\left(gY^2-\frac{g^3}{3}I+4g^2\A^2-[Y,\A]^2\right)
\end{align}
which is precisely the Hoppe model\cite{Hoppe:1982,Hoppe}.

\section{Eigenvalue distribution}
We examine the distribution of eigenvalues of $\X$. I.e. we want to determine the density $\varrho(z_1,\ldots,z_N)$ corresponding to \eqref{partfunc} such that
\begin{align}
    \varrho(z_1,\ldots,z_N)|dz_1|\cdots |dz_N|
\end{align}
is the probability measure on the eigenvalues of $\X$.
Let us consider the action defined by \eqref{partfunc}, $\mathcal Z=\int e^{-N\mathcal{S}}d\X d\A$ where
\begin{align}
    \mathcal S= \Tr\left(\frac{V(\X)-(V(\X))^\dagger}{2i}-(\X\A-\A\X^\dagger)^2\right).
\end{align}
It is invariant under a simultaneous unitary transformation of $\X$ and $\A$. We therefore go to a basis where $\X$ and $\X^\dagger$ are diagonalized. Letting $z_i$ denote the eigenvalues of $\X$, we have
\begin{align}
    \mathcal S=\sum_{i=1}^N\frac{V(z_i)-\overline{V(z_i)}}{2i}+\sum_{i,j=1}^N (z_i-\overline{z_j})(\overline{z_i}-z_j)|\A_{ij}|^2
\end{align}
where $\A_{ij}$ are the matrix elements of $\A$ in the eigenbasis of $\X$. We can perform the Gaussian $d\A$ integral. This gives
\begin{align}
    \varrho=\frac{1}{\tilde{\mathcal{Z}}}\frac{\prod_{i<j}^N(z_i-z_j)(\overline{z_i-z_j})}{\prod_{i,j=1}^N |z_i-\overline{z_j}|}e^{-N\sum_{i=1}^N\left(\frac{V(z_i)-\overline{V(z_i)}}{2i}\right)}
\end{align}
where $\tilde{\mathcal Z}$ is the partition function $\mathcal Z$ times some factors from integrating out the unitary group and the Gaussian integrals, and normalizes the integral of $\varrho$ to 1. Setting $z_j=x_j+ig,V=z^3/3$ gives the standard representation of the Hoppe model.

We recognize the products of eigenvalues as a Cauchy determinant. This leads to a representation of the measure for the eigenvalues as a determinant of the matrix $\mathcal{K}_{ij}=\mathcal{K}(z_i,z_j)$ given by
\begin{align}
    \mathcal{K}(x,y)=\frac{iw(x)\overline{w(y)}}{x-\overline y} \label{kernel}
\end{align}
where $w(x)=\exp(-N V(x)/2i)$. Thus
\begin{align}
    \varrho = \frac{1}{\tilde{\mathcal{Z}}}\det \left[\mathcal{K}(z_i,z_j)\right]_{ij}.
\end{align}
This determinantal representation was written for the Hoppe model case in \cite{Kazakov:1998ji}. A similar determinantal representation of a model closely related to the Hoppe model was observed in \cite{Hoppe:1999xg} and another similar kernel was studied in \cite{zamolodchikov}, corresponding to our case with a single contour being the positive imaginary axis with a different potential.

\section{Saddle point equations}
Another way to write the eigenvalue distribution is as a Gibbs measure
\begin{align}
    \varrho = \frac{1}{\tilde{\mathcal{Z}}}e^{-NH}
\end{align}
where
\begin{align}
    H=\sum_{i=1}^N \frac{V(z_i)-\overline{V(z_i)}}{2i}+\frac{\log|z_i-\overline{z_i}|}{N}-\frac{1}{N}\sum_{i<j}\log\left(\frac{(z_i-z_j)(\overline{z_i-z_j})}{(z_i-\overline{z_j})(\overline{z_i}-z_j)}\right).
\end{align}
This Hamiltonian has a simple interpretation as charged particles interacting with the two-dimensional Coulomb potential $\ln|r|^2$ close to a conducting plate located on the real axis. The denominator is the result of imposing Dirichlet boundary conditions on the real line, which can be interpreted as an infinite conducting plate.

As $N\rightarrow \infty$, the distribution localizes to the global minimum of $H$. Let $\rho(z) |dz|$ be the single-eigenvalue density. The first log-term in $H$ is small and can be dropped. Setting the gradient of $H$ to zero gives the saddle point equations, which can be written on the form
\begin{align}
    \mathrm{Im}\left(\frac{dz_j}{dt}\left(V'(z_j)+2i\int \left(\frac{1}{z'-z_j}-\frac{1}{\overline{z'}-z_j}\right)\rho(z')|dz'| \right)\right)=0,
\end{align}
for $j=1,\ldots,N$,
where $\rho(z)$ is the one-particle density
\begin{align}
    \rho(z)=\left\langle\frac{1}{N}\sum_{i=1}^N \delta(z-z_i)\right\rangle
\end{align}
with $\delta$ taken with respect to $|dz|$, and $dz_j/dt$ is the derivative of $z_j$ along the corresponding trajectory. We recognize the resolvents
\begin{align}
    \int \frac{\rho(z')|dz'|}{z'-z_j}=R_\X(z_j).
\end{align}
Indeed, we have
\begin{align}
    \mathrm{Im}\,\left(W(z)\frac{dz}{dt}\right)=0 \label{impart}
\end{align}
with $W$ defined in \eqref{Wdef}. The above equation is valid for all $z$ in the support of $\rho$. The solution to this {\it Riemann-Hilbert problem} allows us to recover the one-particle density $\rho$ as the discontinuity of $W$ along the cuts on the contours $\gamma_a$. 

\section{Case of a finite number of horizontal lines}\label{sec:4}
We shall now turn to our primary example which is when the spectrum is given by a collection of horizontal lines in the upper half plane $\gamma_a=\{z\in \mathbb C:\mathrm{Im}\,z=g_a\}$, and the cubic potential as in \eqref{cubic}. Then equation \eqref{impart} is simply
\begin{align}
    \mathrm{Im}\, W(z)=0,
\end{align}
for any $z$ in the support of $\rho$. It can be useful to write the density for each contour separately, i.e. $\rho_a(x)$ is the density of eigenvalues with $\mathrm{Im}\, z=g_a$ and real part equal to $x$. We will assume finite support, such that $\rho_a=0$ if $x\not\in [-B_a,B_a]$.

The type of equation we are solving is very similar to saddle point equations arising in related models \cite{Hoppe:1999xg,Eynard:1995zv} and can be solved the standard way by constructing a suitable conformal map.

\begin{figure}
    \centering
    
    \begin{overpic}[width=0.65\textwidth]{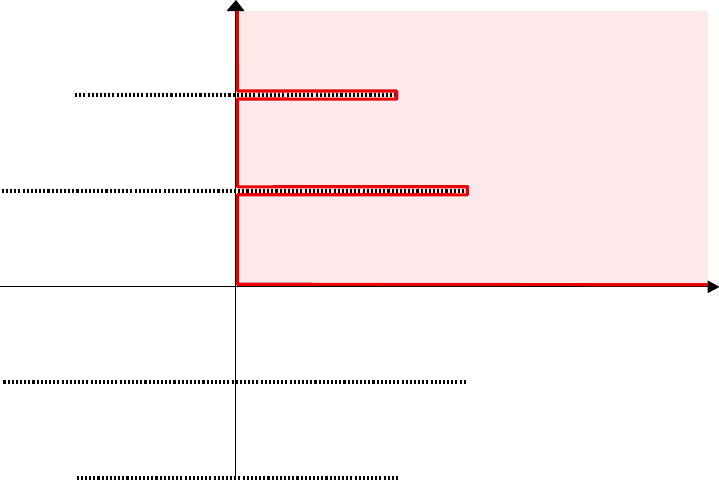}
        \put(102,26){\footnotesize $\mathrm{Re}\, z$}
        \put(30,70){\footnotesize $\mathrm{Im}\, z$}
    \end{overpic}
    \caption{The function $W$ is real on the real and imaginary axes as well as on each side of the cuts. In particular, it is real on the red contour and maps the pink region to the upper half plane. The picture shows the case of two contours in the upper half plane. The small difference between the horizontal segments above and below each of the dotted lines is actually zero, the separation is added only for illustrative purposes.}
    \label{fig:hoppe}
\end{figure}

By symmetry, $W(z)=W(-z)$ is real on both the real and imaginary axes, and is holomorphic outside of the branch cuts on $\pm [ig_a-B_a,ig_a+B_a]$ for some $B_a>0$ (see figure \ref{fig:hoppe}), and real on each side of the branch cuts.
If we in addition impose the condition that the density $\rho$ vanishes at $z=ig_a+B_a$, then there is a unique function with these analytic properties and the asymptotic $W\sim z^2+\mathcal{O}(1/z^2)$ at large $z$. In fact, $W(z)$ is the inverse function of the Schwarz-Christoffel map \cite{Hoppe,Kazakov:1998ji}
\begin{align}
    z=\frac{1}{2}\int_{E_0}^{W(z)} \left\{ \frac{\prod_{a=1}^n(w-S_a)}{\sqrt{(w-E_0)\prod_{i=1}^n(w-E^+_a)(w-E^-_a)}}\right\}dw \label{chsc}
\end{align}
where
\begin{align}
    E_0=W(0),\quad S_a=W(ig_a+B_a),\quad E_{a}^{\pm}=\lim_{x\searrow 0}W(i(g_a\pm x)).
\end{align}
These values are ordered such that $E_0>E_1^->S_1>E_1^+>E_2^->S_2>\cdots >E_n^+$, i.e. $W$ is increasing along the red curve in figure \ref{fig:hoppe} traversed from $+i\infty$ to $+\infty$. There are thus $3n+1$ constants to be determined. There are $2n$ equations that fix the geometry of the contour: $n$ equations fix the vertical distance between the cuts and the real line, and another $n$ equations determine $S_a$ such that the distance above and below the contour is the same, returning it to the real line. One extra equation ensures the correct asymptotic at infinity $W\sim z^2+\mathcal{O}(1/z^2)$. The remaining $n$ degrees of freedom are fixed by the densities in each contour:
\begin{align}
    P_a=\int_{-B_a}^{B_a} \rho_a(x)dx,\quad P_a\ge 0,\quad \sum_{a=1}^n P_a=1
\end{align}
where $P_a$ is the total density on $\gamma_a$ and is fixed in the canonical ensemble.

It is a well known fact that the integral \eqref{chsc} is an Abelian integral on the hyperelliptic curve
\begin{align}
    y^2=Q(W) \label{curve}
\end{align}
where $Q$ is a degree- $2n+1$ polynomial given by
\begin{align}
Q(W)=(W-E_0)\prod_{a=1}^n(W-E_{a}^+)(W-E_a^-) = W^{2n+1}+Q_{2n}W^{2n}+\cdots \label{Qdef}
\end{align}
This defines a Riemann surface of genus $n$. In particular, the degree of $Q$ being odd means it is an imaginary hyperelliptic curve. A contour in the complex plane surrounding a cut of $W(z)$ describes a loop around a handle on the Riemann surface. For the cut in the upper half plane at imaginary part $g_a$, we call this loop the $a$-th $A$-cycle. This allows us to characterize $dz$ as the Abelian differential normalized over the $A$-cycles:
\begin{align}
    \oint_{A_a} dz=0
\end{align}
with a single second-order pole at infinity of the hyperelliptic curve \eqref{curve}.

We use these data to express the free energy of the model
\begin{align}
    F=\frac{1}{N}\langle H(z)\rangle=\frac{1}{N}H(z_*)
\end{align}
by the saddle point approximation. Define the function
\begin{align}
    h_a(x)=g_ax^2-\frac{g_a^3}{3}-\sum_{b=1}^n\int_{-B_b}^{B_b} \log\left(\frac{(x-x')^2+(g_a-g_b)^2}{(x-x')^2+(g_a+g_b)^2}\right) \rho_b(x')dx'. 
\end{align}
Notice that 
\begin{align}
    h_a'(x)=\mathrm{Im}(W(x+ig_a))=0,\quad x\in[-B_a,B_a]
\end{align}
thus, $h_a(x)=h_a$ is a constant for $x\in [-B_a,B_a]$. It is an integral over $B$-cycles of the Abelian differential $W dz$, where the $B$-cycles connect the adjacent points $0,ig_1,ig_2,\ldots,ig_n$ along the imaginary axis in $z$-plane, and goes back on the second sheet where $dz$ takes the opposite sign. Now, the free energy can be expressed as
\begin{align}
    F &= \frac{1}{2}\sum_a\int _{-B_a}^{B_a}\left(g_ax^2-\frac{g_a^3}{3}+h_a\right)\rho_a(x)dx\\
    &=\frac{1}{6Ni}\left\langle\Tr\,\X^3\right\rangle+\frac{1}{2}\sum_a h_a P_a.\label{freeenergy}
\end{align}
The second terms are expressed in terms of the Abelian integral $\int W dz$. The first term is a coefficient in the asymptotic expansion of $W$, as is evident from \eqref{asymptoticW}.

In \eqref{freeenergy} we used $\langle \Tr\, \X^{\dagger 3}\rangle=-\langle \Tr\, \X^3\rangle$.
This is a consequence of the symmetry $z\mapsto-\bar z$. We can write
\begin{align}
    \mathrm{Im}\, \left \langle \Tr \X^{2m}\right \rangle = 0,\quad \mathrm{Re}\,\left\langle \Tr \X^{2m+1}\right \rangle = 0.
\end{align}
We can then write the expansion of $W$ \eqref{asymptoticW} as
\begin{align}
    W\sim z^2 + \frac{4}{iN}\left(\frac{\langle \Tr \X\rangle }{z^2}+\frac{\langle \Tr \X^3\rangle}{z^4}+\cdots\right)\label{momentslines}
\end{align}
By expanding the integral \eqref{chsc} at infinity, we can compute the odd moments. A simple calculation leaves
\begin{align}
    \frac{1}{iN}\langle \Tr\,\X\rangle = \sum_a P_a g_a=-\frac{1}{12}&Q_{2n-1}+\frac{1}{48}Q_{2n}^2+\frac{1}{6}\sum_{a<b}S_aS_b\\
    \begin{split}
    \frac{1}{iN}\langle \Tr\,\X^3\rangle = \frac{1}{20}&\left(-Q_{2n-2}+Q_{2n}Q_{2n-1}-\frac{1}{4}Q_{2n}^3\right.\\&-\left.Q_{2n}\sum_{a<b}S_aS_b-2\sum_{a<b<c}S_aS_bS_c\right)\label{moment3}
    \end{split}
\end{align}
where $Q_j$, defined in equation \eqref{Qdef}, are symmetric polynomials in $E_0,E_a^\pm$, and $E_0,E_a^\pm,S_a$ are defined in \eqref{chsc}.

The odd moments are thus expressible purely in the coefficients of the defining polynomial of the genus-$n$ surface and the "turning points" $S_i$. The even moments seem to be not accessible in the same way. They can, of course, be expressed in terms of integrals of the jump discontinuities of $W$.

Using \eqref{freeenergy} and \eqref{moment3}, we have an expression for the saddle point free energy in terms of the curve data and the $B$-periods of the integral.

\section{Conclusion} By reinterpreting the Hoppe model as describing a normal matrix, we find a generalization. The saddle point equations describe a Coulomb gas with a conducting boundary. When eigenvalues are confined to curves with a fixed distance from the boundary, the solution is a hyperelliptic curve, and the free energy and odd moments are expressed in terms of particular hyperelliptic integrals on the Riemann surface.

A natural question is whether this model plays any distinguished role in the areas where the Hoppe model has previously appeared. This question is currently under investigation. The model does appear in a different context, namely large-$N$ limits of Bethe-ansatz solvable models \cite{mwz}.

Several generalizations are possible. First, this text primarily studies the case where the eigenvalues have an imaginary part belonging to a finite set of positive numbers, but many other choices of subsets of the eigenvalues are possible. For example, one could consider the case of eigenvalues confined to the half circle, corresponding to a subset of unitary matrices, or to the positive imaginary axis with suitable $V$. The latter model has an action in terms of the anti-commutator $(XA-AX^\dagger)^2=\{X,A\}^2$ as in \cite{Eynard:1998he}. 

Second, the kernel $\mathcal{K}(x,y)$ has a natural generalization that preserves the property of being a Cauchy matrix by replacing $1/(x-\bar{y})$ with $1/\sinh (x-\bar{y})$. It can be shown that this leads to a similar saddle point equation which can be solved with the same method. 

\section*{Acknowledgements}
I am deeply grateful to Konstantin Zarembo for suggesting that I study the Hoppe model, reading this text and giving valuable suggestions, Edwin Langmann for reading this text and giving many valuable suggestions, and Paul Wiegmann for many useful discussions and advice. The work was supported by VR grant 2023-04726.

\clearpage
\section*{Appendix: Expansion in moments}
We derive the expression for the first two odd moments from the asymptotic expansion of the integral \eqref{chsc}. At large $z$, the integral \eqref{chsc} is given by
\begin{align}
    z &\sim \sqrt{W}-\frac{1}{2}\int_{W}^\infty\left(\frac{a_0}{w^{3/2}}+\frac{a_1}{w^{5/2}}+\frac{a_2}{w^{7/2}}+\cdots\right)dw\label{asymptote}\\
    &=\sqrt{W}-\frac{a_0}{\sqrt W}-\frac{a_1}{3W^{3/2}}-\frac{a_2}{5W^{5/2}}+\cdots\label{ka}
\end{align}
where $a_i$ are polynomials in $Q_j$ of the polynomial $Q$ defined in \eqref{Qdef} and in $\Sigma_j$, where
\begin{align}
    \Sigma_j=\sum_{a_1<\cdots<a_j}S_{a_1}\cdots S_{a_j}.
\end{align}
The asymptotic expansion \eqref{asymptote} follows from the complete integral
\begin{align}
    \int_{E_0}^\infty \left\{ \frac{\prod_{a=1}^n(w-S_a)}{\sqrt{\prod_{i=1}^n(w-E^+_a)(w-E^-_a)}}-1\right\}\frac{dw}{2\sqrt{w-E_0}}=0
\end{align}
which can be checked by a contour deformation argument just as in \cite{Hoppe} except that now the transformed integral will have several cuts.

The first three terms in the expansion are
\begin{align}
    a_0&=-\frac{1}{2}Q_{2n}-\Sigma_1\\
    a_1&=\frac{1}{2}\left(-Q_{2n-1}+Q_{2n}\left(\frac{3}{4}Q_{2n}+\Sigma_1\right)+2\Sigma_2\right)\\
    \begin{split}
    a_2&=\frac{1}{2}\left(-Q_{2n-2}+\frac{3}{2}Q_{2n}Q_{2n-1}-\frac{5}{8}Q_{2n}^3+\left(Q_{2n-1}-\frac{3}{4}Q_{2n}^2\right)\Sigma_1\right.\\&\left.-Q_{2n}\Sigma_2-2\Sigma_3\vphantom{\frac{1}{2}}\right)
    \end{split}
\end{align}
The asymptotics of $W$ forces $a_0=0$. This simplifies $a_1$ and $a_2$ to be
\begin{align}
    a_1 = \frac{1}{2}\left(-Q_{2n-1}+\frac{1}{4}Q_{2n}^2+2\Sigma_2\right)\\
    a_2=\frac{1}{2}\left(-Q_{2n-2}+Q_{2n}Q_{2n-1}-\frac{1}{4}Q_{2n}^3-Q_{2n}\Sigma_2-2\Sigma_3\right)
\end{align}
Inverting the series \eqref{ka} then gives
\begin{align}
    W(z)\sim z^2+\frac{2a_1}{3z^2}+\frac{2 a_2}{5z^4}+\cdots
\end{align}
and comparing with \eqref{momentslines} we get
\begin{align}
    \frac{1}{N}\langle \Tr\, \X\rangle = \frac{ia_1}{6},\quad \frac{1}{N}\langle \Tr\, \X^3\rangle = \frac{i a_2}{10}
\end{align}
The coefficients have been worked out in the genus 1 case which corresponds to the Hoppe model \cite{Kazakov:1998ji} (notice that they have a different meaning of $g$).

\printbibliography

\end{document}